\documentclass[aps,nofootinbib,preprint]{revtex4}
\usepackage{amsmath,amssymb}
\usepackage{graphicx}

\def\bi{\boldsymbol i}
\def\bq{g}

\begin{document}

\title{
Gauge symmetry breaking in ten-dimensional Yang-Mills theory
dynamically compactified on S$^6$
}

\date{December 16, 2009}

\author{Pravabati Chingangbam\footnote{prava(at)kias.re.kr}}
\affiliation{Korea Institute for Advanced Study, 
207-43 Cheongnyangni 2-dong, Dongdaemun-gu, Seoul 130-722, Republic of Korea}
\author{Hironobu Kihara$^{1,3}$\footnote{hkihara(at)th.phys.titech.ac.jp}}
\author{Muneto Nitta$^{2,3}$\footnote{nitta(at)phys-h.keio.ac.jp}}
\affiliation{$^1$Faculty of Science and Interactive Research Center of Science, 
Graduate School of Science and Engineering,
 Tokyo Institute of Technology, 2-12-1 Oh-okayama, Meguro, Tokyo 152-8551, Japan\\
\it $^2$Department of Physics, and 
$^3$Research and Education Center for Natural Sciences,
Keio University,\\ 
4-1-1 Hiyoshi, Yokohama,
Kanagawa 223-8521, Japan
}

\preprint{KIAS-P09057}
\preprint{TIT/HEP-601}

\begin{abstract}
We study fluctuation modes in ten-dimensional Yang-Mills theory with a 
higher derivative term for the gauge field. 
We consider the ten-dimensional space-time to be a product of a four-dimensional 
space-time and six-dimensional sphere which exhibits dynamical compactification. 
Because of the isometry on $S^6$, 
there are flat directions corresponding to the 
 Nambu-Goldstone zero modes in the effective theory on the solution. 
The zero modes are absorbed into gauge fields and
form massive vector fields as a consequence of the Higgs-Kibble 
mechanism.
The mass of the vector fields is proportional to the inverse of the
radius of the sphere and larger than the mass scale
set by the radius because of the
higher derivative term. 
\end{abstract}

\maketitle
\section{Introduction}
The idea of extra dimensions and their compactification 
have been attracting people \cite{Nordstrom:1988fi} 
for several decades. 
Among many scenarios of compactification of extra dimensions, 
Cremmer and Scherk suggested an interesting idea 
that compactification may occur if a topologically 
non-trivial gauge configuration exists in 
compactified space \cite{Cremmer:1976ir}. 
An example is given by 't Hooft-Polyakov monopole \cite{Polyakov:1974ek} 
on $S^2$.
Recently in \cite{Kihara:2009ea} 
some of us have studied a scenario of dynamical compactification
and inflation in ten-dimensional Einstein-Yang-Mills theory 
with SO(6) gauge group, 
using the Cremmer-Scherk gauge configuration 
on $S^6$ \cite{Kihara:2007di,Kihara:2007vz}. 
We have added a higher derivative coupling term,\footnote{  
Such a quartic term is known to
appear in the low-energy effective theory of quantum electro dynamics 
too \cite{Heisenberg:1935qt}.
}  
originally introduced by Tchrakian \cite{Tchrakian:1978sf},  
in order to ensure the non-existence of tachyonic modes
on the Cremmer-Scherk configuration because of the Bogomol'nyi equation. 

In this paper we study the issue of the stability and fluctuations 
of the Cremmer-Scherk configuration in this scenario. 
We will work on the space-time given by the product of 
a four-dimensional space-time and 
a six-dimensional sphere, 
where the radius of the sphere shrinks to a constant value 
in the limit 
$t \rightarrow + \infty$. 
We assume that the four-dimensional space-time 
can be treated as $\{ (t, {\cal N}_t)\}_{t \in {\mathbb R}}$, where
${\cal N}_t$ is diffeomorphic to a 
three-dimensional manifold ${\cal N}$.\footnote{For instance, 
four-dimensional Friedmann-Lemaitre-Robertson-Walker space-time 
satisfies the condition.} 

We first show the absence of tachyonic modes in 
general background gauge fields, 
and then study if there are massless modes or not.
In general
the Cremmer-Scherk configuration is obtained by the identification of compact
direction and internal (gauge) direction 
as an extension of 't Hooft-Polyakov monopole. 
By rotating this identification, with  
the rotation depending on ${\mathbb R}^{1,3}$, 
we obtain massless fluctuation modes which 
can be regarded as Nambu-Goldstone 
bosons \cite{Nambu:1961tp}. 
Then we show that these massless modes are actually 
absorbed into gauge fields to form massive vector 
(Proca) fields 
by the Higgs-Kibble mechanism \cite{Higgs:1964pj}.
Since the number of Nambu-Goldstone modes, 
fifteen, coincides with the dimension of SO(6), 
we find that 
the gauge symmetry SO(6) is completely broken. 
By scaling fields for
normalization of the coefficient of their kinetic terms, we obtain
the mass proportional to the inverse of the radius of the compact space. 
We thus conclude that there are neither tachyonic nor 
massless modes in the physical spectrum around the background 
and that the configuration is stable.
 Similar mechanisms for generating masses in compactifications 
have been discussed 
in the literature.
Scherk and Schwarz introduced mass 
by a generalized dimensional reduction or 
a twisted boundary condition of fields \cite{Scherk:1979zr}. 
Horvath {\it et. al.} considered
system coupled with Higgs fields to obtain mass of gauge fields  
\cite{Horvath:1977st}. 
Manton showed that the components of gauge 
fields along extra-dimensions provide Higgs fields in the
four dimensional effective theory without additional scalar fields
\cite{Manton:1979kb,Forgacs:1979zs}.  
Hosotani found a mechanism for obtaining Higgs fields belonging to 
representations different from the adjoint representation 
and endowing mass to fermions
on orbifold \cite{Hosotani:1983xw}. 

This paper is organized as follows:  
in Sec. \ref{section:genfluc} we will study the general treatment of
the fluctuation of gauge fields about some classical solution. 
In the effective theory on the classical background,
the fluctuation of the ten-dimensional vector field 
splits
into two
parts, a four-dimensional vector field and six scalar fields.
We will see that the lowest Kaluza-Klein mode of the 
four-dimensional vector field gets mass because of the gauge 
configuration. In Sec. \ref{section:csandt} we review the 
Cremmer-Scherk gauge configuration and the Tchrakian's self-duality 
equation.  In Sec. \ref{section:njg} we show the Higgs mechanism
during the dynamical compactification. By considering rotated
identification between the internal (gauge) space and the compact space
($S^6$), 
we explicitly show that the Nambu-Goldstone bosons
form the Proca fields with vector fields.
In Sec. \ref{section:conclusion} we  summarize this article.  
In Appendix A, we prove the non-existence of covariantly
  constant functions belonging  
to the adjoint representation on the Cremmer-Scherk configuration. 

\section{Fluctuations around solutions of Bogomol'nyi equation}
\label{section:genfluc}

Let us consider the SO(6) Yang-Mills theory on the ten-dimensional 
curved space-time which is a
direct product of a four-dimensional space-time ${\cal M}$ and a 
six-dimensional sphere $S^6$, whose radius varies.  
We assume that the four-dimensional space-time ${\cal M}$
is the form of ${\cal M}= \{ (t, {\cal N}_t)\}_{t \in {\mathbb R}}$, where
${\cal N}_t$ is diffeomorphic to a three-dimensional space ${\cal N}$, 
and that any tangent vectors on the time slice ${\cal N}_t$ are space like.
The metric on the ten-dimensional space is given as 
\begin{align}
ds^2 &= g_{\mu\nu} (x) dx^{\mu} dx^{\nu} + g_{ij}(x,y) dy^i dy^j~,& 
g_{ij} (x,y) &= R_2(x)^2 \frac{\delta_{ij}}{(1+|y|^2/4)^2} ~.
\end{align}
The indices are $\mu, \nu =  0 , 7, 8 , 9 $ and $i,j =  1, \cdots, 6$. 
$x^{\mu}$ represent four-dimensional coordinates along 
the four-dimensional space-time ${\cal M}$ 
and $y^i$ represent 
six-dimensional coordinates along the sphere. 
We denote $X^M\equiv(x^{\mu},y^i)$ for the coordinates of total 
space-time and their indices are represented by capital scripts.  
$R_2(x)$ is the radius of the six-dimensional sphere and depends on
 the four-dimensional coordinates $x$. 
We consider only the case where $R_2(x)$ converges 
to a nonzero constant value in the limit $t \rightarrow + \infty$. 
We start from the following action, 
\begin{align}
S &:= \frac{1}{16} \int {\rm Tr} \left\{  - F \wedge * F 
+ \alpha^2 (F \wedge F) \wedge * (F \wedge F)  \right\} ~,
 \label{eq:action}
\end{align}
where $F$ is the field strength two-form and $\alpha$ is the quartic
coupling constant. 
The second term quartic in $F$ is the term introduced by Tchrakian 
\cite{Tchrakian:1978sf} which we call the Tchrakian term. 
This action is quadratic in the time derivative acting on the gauge field, $\partial A/\partial t$.
We consider the gauge potential one-form $A$ taking 
a value in so(6). 
In our notation
the generators of so(6) are represented with spinor indices,   
but one should not confuse 
them with spinor fields.  
In order to define the product of field strength two-form, 
we have to indicate the representation matrix of gauge fields. 
Let us use the Clifford algebra $\{ \gamma_a , \gamma_b \} = 2
\delta_{ab}$,  $(a,b=1,2,\cdots, 6)$ with respect to SO(6). 
Here $\gamma_a$ are Hermitian $8 \times 8$ matrices. 
The internal indices are represented by $a,b,c, \cdots$. 
This is independent of the space indices along the six-dimensional sphere. 
Commutators $\gamma_{ab}:= (1/2) [ \gamma_a , \gamma_b ]$ of
$\gamma$-matrices satisfy the commutation relation of so(6) and their normalization is shown as follows,
\begin{align}
[\gamma_{ab} , \gamma_{cd} ] &=2 ( \delta_{bc} \gamma_{ad} - \delta_{bd} \gamma_{ac} - \delta_{ac} \gamma_{bd} + \delta_{ad} \gamma_{bc} ) ~,\cr
{\rm Tr} \gamma_{ab} \gamma_{cd} &= 8 ( \delta_{bc} \delta_{ad} - \delta_{ac}\delta_{bd}  )~.
\end{align}
The anticommutation relation of these generators $\gamma_{ab}$ is
given by
\begin{align}
\{ \gamma_{ab} , \gamma_{cd}  \} &=  2 \gamma_{abcd}  + 2 (
\delta_{bc} \delta_{ad} - \delta_{ac} \delta_{bd})~, 
\end{align}
where we have used the following notation,
\begin{align}
\gamma_{a(1) \cdots a(p)} & := \frac{1}{p!} \sum_{\sigma \in S_p} {\rm sgn} ( \sigma ) 
\gamma_{a(\sigma(1)) }  \gamma_{a(\sigma(2)) } \cdots \gamma_{a(\sigma(p)) } ~,
\end{align}
where $S_p$ is the symmetric group of $p$ characters. 
Further, we use the matrix $\gamma_7 := -\bi \gamma_{123456}$. 
The gauge potential can then be written as
\begin{align}
A &:= \frac{1}{2} A_M^{ab} \gamma_{ab} dX^{M}~.
\end{align}
The field strength is defined as $F := dA + \bq A \wedge A$,  
with gauge coupling constant $\bq$.  
The equation of motion 
of the Lagrangian (\ref{eq:action}) reads
\begin{align}
D\left\{ * F  - \alpha^2 \left(  F \wedge * (F \wedge F)   +
(F \wedge F ) \wedge F \right) \right\} &=0~. 
\label{eq:eom}
\end{align}
Let us suppose that $A^{(0)}$ is some solution of 
Eq. (\ref{eq:eom}), which
we fix to be our background solution. We make
the assumtions that $A^{(0)}$   
has components only along the compact directions and depends only on $y^i$. 
Then it follows that the corresponding field strength, $F^{(0)} := d A^{(0)} +
\bq A^{(0)} \wedge A^{(0)}$,  has components only along the compact directions. 

Let us consider fluctuations $\delta A$ around 
the solution $A^{(0)}$, as 
\begin{align}
A &= A^{(0)} + \delta A~,& 
\delta A &= v + \Phi~,&
v &= v_{\mu} dx^{\mu}~,& 
\Phi &= \Phi_i dy^i  \label{eq:fluc}
\end{align}
The fluctuation $\delta A$  is divided into two parts, $v$  
and $\Phi$. 
$v$ is a one-form whose components are nonzero only along the four-dimensional space-time, 
while $\Phi$ has components only along the
six-dimensional sphere. 
The coefficients depend on $x$ and $y$, $v_{\mu} := v_{\mu}(x,y)$, 
$\Phi_{i} := \Phi_{i}(x,y)$.  
Our objective is then to obtain the four-dimensional effective theory for these
fluctuations. 
The fluctuation $v$ is  a vector field and 
the fluctuation $\Phi_i, (i=1,2, \cdots , 6)$,  are six scalar fields
 under the general coordinate transformation of the four-dimensional space-time. 
Each $v_{\mu}$ or $\Phi_i$ belongs to the adjoint representation of the gauge group SO(6) and 
$(\Phi_1^{ab} , \Phi_2^{ab} , \cdots , \Phi_6^{ab})$ transform as vectors under the 
rotation of the six-dimensional space. 
Let us decompose the field strengths in terms of $v$ and $ \Phi$, 
\begin{align}
F &= F^{(0)} +  d \delta A + {\bq} 
\left(  A^{(0)}  \wedge \delta A + \delta A \wedge  A^{(0)}  
\right) +  {\bq} 
 \delta A \wedge \delta A~\cr
&= F^{(0)} +  d_{(4)} v  + d_{(4)} \Phi 
+  d_{(6)} v  + d_{(6)} \Phi  \cr
&+   {\bq} 
\left(  A^{(0)}  \wedge v + v \wedge  A^{(0)}  \right)
+  {\bq} 
\left(  A^{(0)}  \wedge \Phi + \Phi \wedge  A^{(0)}  \right)\cr
& +  {\bq}   \left( v \wedge v  + v \wedge \Phi + \Phi 
\wedge v + \Phi \wedge \Phi
\right)
\end{align}
where the exterior derivative $d$ is decomposed into two parts, 
$d= d_{(4)} + d_{(6)}$. 
These operators are defined as 
\begin{align}
d_{(4)} &:=  dx^{\mu} \frac{\partial}{\partial x^{\mu}} ~,& 
d_{(6)} &:=  dy^{i} \frac{\partial}{\partial y^{i}} ~. 
\end{align}
Let us gather terms as 
\begin{align}
W &:= d_{(4)} v + \bq v \wedge v~,\cr
D_v^{(4)} \Phi &:= d_{(4)} \Phi + \bq ( v \wedge \Phi + 
\Phi \wedge v  )~,\cr
 D_0^{(6)}  v &:= d_{(6)} v + \bq ( A^{(0)} \wedge v +  v  
\wedge A^{(0)}  )~,\cr
D_0^{(6)} \Phi & := d_{(6)} \Phi + \bq ( A^{(0)} \wedge 
\Phi + \Phi \wedge A^{(0)} ) ~,
\end{align}
where $W$ is a field strength-like quantity of the vector field $v$ 
in four-dimensional space-time. 
$\Phi$ becomes a scalar field which belongs to the adjoint 
representation of SO(6) in four dimensional low energy effective theory. 
$D_v^{(4)}$ represents the covariant exterior derivative with 
 vector field $v$ on the four-dimensional space-time, 
while 
$ D_0^{(6)}$ represents the covariant exterior derivative with classical 
gauge configuration $A^{(0)}$,   
with derivatives along the six-dimensional sphere.  
By using these definitions the field strength, $F$, is written as
\begin{align}
F &= F^{(0)} +  W  + D_v^{(4)} \Phi 
+  D_0^{(6)}  v + D_0^{(6)} \Phi   +  {\bq}   
\left( \Phi \wedge \Phi \right) \cr
&=F_{\rm (c)} + F_{\rm (nc)}~,\cr
F_{\rm (c)} &:= F^{(0)} 
+ D_0^{(6)} \Phi   +  {\bq}   \left( \Phi \wedge 
\Phi \right) \cr
F_{\rm (nc)} &:=  W  + D_v^{(4)} \Phi 
+  D_0^{(6)}  v ~. \label{eq:F,Fc,Fnc}
\end{align} 
We have separated the field strength into two parts according to      
their indices. 
$F_{\rm (c)}$ has components only along compact directions, while 
$F_{\rm (nc)}$ has components along four-dimensional space-time. 
The Yang-Mills part of the action is written as
\begin{align}
{\rm Tr} F \wedge * F &= {\rm Tr} \left( F_{\rm (c)} \wedge * F_{\rm (c)} + F_{\rm (nc)} 
\wedge * F_{\rm (nc)}  \right) ~.
\end{align}
There are no terms obtained by the contraction of $F_{\rm (c)}$ and $F_{\rm (nc)}$. 
Similarly, the higher derivative coupling term is decomposed as
\begin{align}
{\rm Tr} (F \wedge F) \wedge * (F \wedge F) 
&= {\rm Tr} \left[   (F_{\rm (c)} \wedge F_{\rm (c)} ) \wedge * ( F_{\rm (c)} \wedge F_{\rm (c)})
    \right.\cr 
&+  (F_{\rm (c)} \wedge F_{\rm (nc)} + F_{\rm (nc)} \wedge F_{\rm (c)} ) \wedge * ( F_{\rm (c)} \wedge F_{\rm (nc)} + F_{\rm (nc)}
    \wedge F_{\rm (c)} )   \cr
&+
\left. (F_{\rm (nc)} \wedge F_{\rm (nc)} ) \wedge * ( F_{\rm (nc)} \wedge F_{\rm (nc)} )  \right] ~.
\end{align}
We now decompose the action $S$ given in Eq.~(\ref{eq:action}) as
\begin{align}
S & := S_{\rm (c)} + S_{\rm (nc)}~,\cr
S_{\rm (c)}&:=  \frac{1}{16} \int  {\rm Tr} \left[ - F_{\rm (c)} \wedge * F_{\rm (c)} +
  \alpha^2  ( F_{\rm (c)} \wedge F_{\rm (c)} ) \wedge * ( F_{\rm (c)} \wedge F_{\rm (c)} )   \right]
  \cr 
S_{\rm (nc)} &= \frac{1}{16} \int   {\rm Tr} \left[ - F_{\rm (nc)}
  \wedge * F_{\rm (nc)} \right. \cr
&  \left. +   \alpha^2   (F_{\rm (c)} \wedge F_{\rm (nc)} + F_{\rm (nc)} \wedge F_{\rm (c)} ) \wedge * ( F_{\rm (c)} \wedge
  F_{\rm (nc)} + F_{\rm (nc)}  \wedge F_{\rm (c)} ) \right.   \cr
& \left. + \alpha^2  (F_{\rm (nc)} \wedge F_{\rm (nc)} ) \wedge * ( F_{\rm (nc)} \wedge F_{\rm (nc)} )
  \right]  
\end{align}
It was shown in \cite{Kihara:2009ea} that there are no tachyonic modes obtained from $\Phi$ 
in $S_{\rm (c)}$ if the solution $A^{(0)}$ is a solution of the Bogomol'nyi
equation 
given in Eqs.~(\ref{eq:bog}) and (\ref{eq:bogeq}), below.
Now we focus only on the $v$ fluctuations and 
we need to check only the remaining part $S_{\rm (nc)}$. 
In order to study the mass spectrum, 
we need only quadratic terms in $S_{\rm (nc)}$ and all 
such terms are included in the following,  
\begin{align}
S_{\rm (nc)} |_{{\rm quad.}} &=\frac{1}{16} \int   {\rm Tr} \left[ - F_{\rm (nc)}
  \wedge * F_{\rm (nc)}  \right. \cr 
&  \left. + \alpha^2  (F^{(0)} \wedge F_{\rm (nc)} + F_{\rm (nc)} \wedge F^{(0)}  )
  \wedge * ( F^{(0)}  \wedge F_{\rm (nc)} + F_{\rm (nc)} \wedge F^{(0)} )  \right]  ~. 
\end{align}

Let us decompose $F_{\rm (nc)}$ as,
\begin{align}
 F_{\rm (nc)} &= W + F_{\rm (m)}~,&
 F_{\rm (m)} & :=  D_v^{(4)} \Phi +  D_0^{(6)}  v ,
\end{align}
where $W$ has two four-dimensional indices and $F_{\rm (m)}$ has one
four-dimensional index and one six-dimensional index. Therefore, in
this part, $S_{\rm (nc)} |_{{\rm quad.}}$, of this action, there are no
terms obtained by the contraction of $W$ and $F_{\rm (m)}$: 
\begin{align}
S_{\rm (nc)} |_{{\rm quad.}} &=\frac{1}{16} \int   {\rm Tr} \left[ - W
  \wedge * W + \alpha^2  (F^{(0)} \wedge W + W \wedge F^{(0)}  )
  \wedge * ( F^{(0)}  \wedge W + W \wedge F^{(0)} )  \right]  \cr 
&+ \frac{1}{16} \int   {\rm Tr} \left[ - F_{\rm (m)} \wedge * F_{\rm
  (m)} \right. \cr
& \left. + \alpha^2
  (F^{(0)} \wedge F_{\rm (m)} + F_{\rm (m)} \wedge F^{(0)}  ) \wedge * ( F^{(0)}
  \wedge F_{\rm (m)} + F_{\rm (m)} \wedge F^{(0)} )  \right]  ~. 
\label{eq:effkin}
\end{align}
From the second integral of Eq. (\ref{eq:effkin}) we obtain the mass term in
four-dimensional effective theory of $v$ as  
\begin{align}
S_{\rm (nc)} |_{{\rm mass~of~}v } &=  \frac{1}{16} \int   {\rm Tr} \left[
  -  D_0^{(6)}  v   \wedge * D_0^{(6)}  v  + \alpha^2   \{ F^{(0)} ,
  D_0^{(6)}  v  \} \wedge * \{ F^{(0)} , D_0^{(6)}  v  \}  \right]   ~.
\label{eq:massterm}
\end{align}
This gives a mass matrix defined as eigen values of 
a second order differential operator on $S^6$. 
In order to 
show that the square of mass is positive, let us do the
following: suppose that $\omega$ and $\lambda$ are functions or
0-forms on $S^6$ taking values in the Lie algebra so(6). 
The inner product is defined as 
\begin{align}
\langle \omega , \lambda \rangle_6 &:=
- \frac{1}{8} \int_{S^6}  {\rm Tr} \omega \wedge *_6 \lambda 
=- \frac{1}{8} \int_{S^6} d^{(6)} v {\rm Tr} \omega \lambda  ~.
\end{align}
This gives a positive definite norm. 
Any arbitrary function $f$ globally defined on $S^6$
has finite norm. 
$L^2(S^6)\otimes so(6)$ with respect to this norm is a Hilbert space and separable.  
The mass matrix ${\cal M}$ is defined as 
\begin{align}
{\cal M} \lambda  &:=    D_0^{(6)} *_6 D_0^{(6)}  \lambda  - \alpha^2   \{ F^{(0)} ,
 D_0^{(6)} *_6 \{ F^{(0)} , D_0^{(6)}  \lambda  \} \} ~.
\end{align}
This matrix is self-adjoint operator in $L^2(S^6)\otimes so(6)$. 
By using these, we obtain
\begin{align}
\langle \omega , {\cal M} \omega \rangle_6 &= 
\frac{1}{8} \int   {\rm Tr} \left[
  -  D_0^{(6)}  \omega    \wedge * D_0^{(6)}  \omega   + \alpha^2   \{ F^{(0)} ,
  D_0^{(6)}  \omega   \} \wedge * \{ F^{(0)} , D_0^{(6)}  \omega   \}  \right] ~.
\end{align}
The integrand cannot be negative. 
Hence the operator ${\cal M}$ has non-negative eigenvalues. 
Actually this operator can be considered as the mass matrix of $v$. 
The mass term of $v$ is written as
\begin{align}
S_{\rm (nc)} |_{{\rm mass~of~}v } &=  
 \frac{1}{2} \int d^{(4)}v  \langle v_{\mu} , {\cal M} v^{\mu} \rangle_6 ~.
\end{align}
From this we obtain that the vector field $v$ has non-negative 
 square of mass. 
This ensures that there are no tachyonic modes in the
full fluctuation of the gauge field
 around the solution of the Bogomol'nyi equation  
in~\cite{Kihara:2009ea}. 
Apparently, solutions of the equation $D_0^{(6)} v=0$ are massless modes 
if they exist 
although it 
does not mean that only solutions of $D_0^{(6)} v=0$ 
are massless modes. We show the non-existence of covariantly constant 
functions on
the Cremmer-Scherk configuration in Appendix.

\section{Cremmer-Scherk Gauge Configuration and Tchrakian's Duality Equation
} 
\label{section:csandt}
So far we have not identified background configuration. 
Let us now focus on the Cremmer-Scherk gauge configuration 
on $S^6$.
To describe the background solution, 
here let us restrict our interest only to the part $S_{\rm (c)}$  
\begin{align}
S_{\rm (c)}&= -    \frac{1}{16}    \int \sqrt{-g^{(4)}} d^4x \int {\rm Tr} \left\{ - F_{\rm (c)}
\wedge *_6 F_{\rm (c)} + \alpha^2 ( F_{\rm (c)} \wedge F_{\rm (c)} ) \wedge *_6 ( F_{\rm (c)} \wedge
F_{\rm (c)} )    \right\} ~, 
\end{align}
where $F_{\rm (c)}$  has only components along the compact direction and the
Hodge dual  of $F_0$ and $F_0 \wedge F_0$  are split into 
four-dimensional  invariant volume form and the Hodge dual of those on
$S^6$.  This part of the action is a functional of only $\Phi$ and $R_2$. 
Let us define the following quantity 
\begin{align}
M_{\rm (c)}[\Phi,R_2]&:= \frac{1}{16}   \int  {\rm Tr} \left\{ - F_{\rm (c)} \wedge *_6 F_{\rm (c)}
+  \alpha^2 ( F_{\rm (c)} \wedge F_{\rm (c)} ) \wedge *_6 ( F_{\rm (c)} \wedge F_{\rm (c)} )   \right\}  ~.
\end{align}
By using this, the part of the action, $S_{\rm (c)}$, is rewritten as
\begin{align}
S_{\rm (c)}&= - \int \sqrt{-g^{(4)}} d^4x M_{\rm (c)}[\Phi,R_2]~.
\end{align}
From this expression, the term $M_{\rm (c)}[\Phi,R_2]$ seems to be 
a part of the effective action including coupling terms of $R_2$ and $\Phi$. 

The term $M_{\rm (c)}[\Phi,R_2]$ can be treated as pseudo-energy on a space with Euclidean signature. 
By the Bogomol'nyi completion, it can be rewritten as 
\begin{align}
M_{\rm (c)}[\Phi,R_2]&= - \frac{1}{16}   \int {\rm Tr} \left\{ F_{\rm (c)} \mp *_6 \bi
\alpha   \gamma_7 F_{\rm (c)}  \wedge F_{\rm (c)}  \right\} 
\wedge *_6  \left\{ F_{\rm (c)}  \mp *_6 \bi  \alpha  \gamma_7 F_{\rm (c)}  \wedge F_{\rm (c)}
\right\}  \cr
& \mp \frac{\bi}{8} \alpha \int {\rm Tr} \gamma_7 F_{\rm (c)}  \wedge F_{\rm (c)}  \wedge F_{\rm (c)}  ~, \label{eq:bog}
\end{align}
and the Bogomol'nyi equation is obtained as
\cite{Kihara:2007di,Kihara:2007vz}
\begin{align}
F_{\rm (c)}  \mp *_6 \bi  \alpha  \gamma_7 F_{\rm (c)}  \wedge F_{\rm (c)}   &=0~.
\label{eq:bogeq}
\end{align}
The term   
\begin{align}
 Q: = {\rm Tr} \gamma_7 F_{\rm (c)}  \wedge F_{\rm (c)}  \wedge F_{\rm (c)}  
\end{align} 
is a total derivative and the integral over it reduces to a surface integral. 
The resultant of the integral gives topological quantity and 
$M_{\rm (c)}[\Phi,R_2]$ is  bounded from below. 
The solution of Eq. (\ref{eq:bogeq}) with the minimal 
topological charge has been given  
in \cite{Tchrakian:1978sf,Kihara:2007di,Kihara:2007vz}.
The minimal charge is 
\begin{align}
Q = \frac{96 \pi^3}{\bq^3} ~. \label{eq:charge1}
\end{align}

The gauge configuration 
\begin{align}
A^{(0)} = \frac{1}{ 4 \bq R_2} \gamma_{ab} y^{a} V^{b} ~, \quad
F^{(0)} = \frac{1}{ 4 \bq R_2^2} \gamma_{ab} V^{a} \wedge V^{b} 
 \label{eq:sol-R}
\end{align}
satisfies the ``self-duality" equation 
\begin{align}
F^{(0)} &=  *_6 \bi  \frac{\bq R_2^2}{3}  
\gamma_7 F^{(0)} \wedge F^{(0)}  ~,
\label{eq:sde}
\end{align} 
where $R_2$ is the radius of $S^6$ and 
$V^{i}$ are vielbein of $S^6$, given as 
\begin{align}
 V^{i} &= R_2 \frac{dy^i}{(1+|y|^2/4)} .~
\end{align} 
The configuration (\ref{eq:sol-R}) solves 
the Bogomol'nyi equation (\ref{eq:bogeq}) 
if and only if 
$R_2$ takes the special constant value determined by the constant $\alpha$   
\cite{Kihara:2007vz}, given as  
\begin{equation}
\alpha = \bq R_2^2/3. 
\end{equation}
Let us call this special radius $L_c := \sqrt{3\alpha/\bq}$ as the
``{\it Bogomol'nyi radius}".  
In the case of $\alpha \neq \bq R_2^2/3$, 
the gauge configuration (\ref{eq:sol-R}) do not 
satisfy the Bogomol'nyi equation (\ref{eq:bogeq}) anymore.  

In this article, we are considering the metric with 
$R_2(x) \rightarrow L_c$ in the limit $t \rightarrow + \infty$.
An example of such a realization is given
in~\cite{Kihara:2009ea}, where the four-dimensional part $g_{\mu\nu}$
 is given by the Friedmann-Lemaitre-Robertson-Walker metric. 
The gauge field configuration then approaches the 
solution of
Bogomol'nyi equation. When $R_2(x)$ is close to $L_c$, we can consider
perturbation of $R_2(x)$ about $L_c$, such that the deviation is
quantified by a scalar field, $\phi_2(x)$:
\begin{align}
R_2 &= L_c \exp(\phi_2(x))~.
\end{align}
  $L_c$ being a stable fixed
point ensures that  $\phi_2(x)$ approaches zero at all spatial points
of (1,3) part of space-time. Note that in~\cite{Kihara:2009ea} we had
considered only 
the case of
spatially constant $\phi_2(x)\equiv \phi_2(t)$ case
and shown that it approaches zero with time. 

We are now ready to study mass spectrum around the background.
To this end, 
let us first expand $M_{\rm (c)}$ as
\begin{align}
M_{\rm (c)}[\Phi,R_2]|_{\rm \Phi^2} &= M_{\rm (c)}[\Phi,R_2]|_{\rm \Phi^2}^{(1)}  + M_{\rm (c)}[\Phi,R_2]|_{\rm \Phi^2}^{(2)} 
\end{align}
with
\begin{align}
M_{\rm (c)}[\Phi,R_2]|_{\rm \Phi^2}^{(1)} 
&= - \frac{1}{16} \int {\rm Tr} 
\left\{ D_0^{(6)} \Phi - \bi *_6 \alpha \gamma_7 \{ F^{(0)} , D_0^{(6)} \Phi \}   \right\}  \cr
& \wedge *_6 \left\{ D_0^{(6)} \Phi - \bi *_6  \alpha \gamma_7 \{ F^{(0)} , D_0^{(6)} \Phi \}   \right\}
\\  
M_{\rm (c)}[\Phi,R_2]|_{\rm \Phi^2}^{(2)} 
& = - \frac{\bi}{16} \frac{\bq R_2^2}{3} \left( 1 +
  e^{-4 \phi_2(x)} \right) \int 
{\rm Tr} \gamma_7 F^{(0)} \wedge F^{(0)} \wedge F^{(0)} 
  \cr
&- \frac{\bi \bq}{8} \frac{\bq R_2^2}{3}  \int {\rm Tr} \left( 1 -
e^{-2 \phi_2(x)}  \right)^2  
      \gamma_7 F^{(0)} \wedge F^{(0)} \wedge \left( \Phi \wedge 
\Phi \right)   \cr
& +{\rm total~ derivative} ~.
\end{align}
The first term $M_{\rm (c)}[\Phi,R_2]|_{\rm \Phi^2}^{(1)}$ 
has only positive eigenvalues in the mass matrix, 
and so it gives a positive contribution to 
the eigenvalues of the total mass matrix. 
On the other hand the second term 
$M_{\rm (c)}[\Phi,R_2]|_{\rm \Phi^2}^{(2)}$ 
includes terms which lower the eigenvalues of the mass matrix. 
This term
vanishes for $\phi_2 = 0$,
which is realized in dynamical compactification \cite{Kihara:2009ea}. 
%
%
As we show later there are massless modes, satisfying 
\begin{align}
D_0^{(6)} \Phi - \bi *_6 \alpha \gamma_7 \{ F^{(0)} , D_0^{(6)} \Phi \}  &=0~.
\end{align}
This equation is satisfied for the deformation of gauge configuration, 
after which the compact part $F_{(c)}$ of the deformed configuration still
satisfies the Bogomol'nyi equation. 
Because these zero modes are not included in 
$M_{\rm (c)}[\Phi,R_2]|_{\rm \Phi^2}^{(2)}$  as shown 
in Eqs. (\ref{eqn:act1}) and (\ref{eq:value}) below, 
we do not care about the tachyonic modes coming from these zero modes. 
Hence the eigenvalue of the lowest modes 
which give non-zero contribution to $M_{\rm (c)}[\Phi,R_2]|_{\rm \Phi^2}^{(1)}$
gives a gap. This implies that there is a range of $\phi_2$ 
in which the mass matrix does not have tachyonic eigenvalues. 

Around 
this background the vector field, $v_{\mu}(x)$, becomes massive and
the mass term is  
\begin{align}
S_{\rm mass}
 &= - \frac{16 \pi^3}{15} L_c^4 \int  \sqrt{-g^{(4)}} d^4x         \left( 1 +  \frac{10}{9} e^{-4 \phi_2}   \right) v_{\mu}^{ab} v^{\mu,ab} ~.
\end{align}
Since $S_{\rm mass}$ is quadratic in $v$ and our interest is in
getting the mass terms, we replace $R_2$ by  $L_c$. 
This implies that the gauge symmetry is broken completely. 
Hence, we expect that the Higgs-Kibble mechanism occurs on this
background. We show it explicitly in the next section.

\section{Rotation of Cremmer-Scherk Configuration and Higgs mechanism}
\label{section:njg}
The gauge configuration is obtained by the identification of compact
directions and internal directions.  
Here we consider the rotated identification. 
Let us use the following quantities, 
\begin{align}
z^a &= U^{ai} y^i~,& W^a &= U^{ai} V^i~,&
U^{ai} U^{bi}  &= \delta^{ab}~,& U^{ci} U^{cj}  = \delta^{ij}~,
\end{align}
where the rotation matrix $U^{ai} \in$ SO(6) 
has different types of indices. 
Note that this is different from gauge transformations in general.  

When $U^{ai}$ is a constant matrix  
the rotation can be 
absorbed into gauge transformations. 
As a first step, let us see what happens to 
the Cremmer-Scherk gauge configuration under such a constant rotation.
$W^a$ as well as $V^i$ form vielbein as 
\begin{align}
ds^2 &= \delta_{ij} V^i V^j = \delta_{ij} W^i W^j~.
\end{align}
Therefore, the Hodge dual of the four-product of $W$ is given as
\begin{align}
*_6 W^{ijkl} &= \frac{1}{2} \epsilon^{ijklmn} W^{mn}~.
\end{align}
Then the gauge configuration  
\begin{align}
A^{(U)} &:=  \frac{1}{ 4 \bq R_2} \gamma_{ab} z^{a} W^{b} ~,& 
F^{(U)} &= 
d A^{(U)}  + \bq A^{(U)}  \wedge A^{(U)} =
\frac{1}{ 4 \bq R_2^2} \gamma_{ab} W^{a} \wedge W^{b} ~,
\end{align}
satisfies the self-duality equation, 
 \begin{align}
F^{(U)}   &= *_6  \frac{\bq R_2^2}{3} \bi \gamma_7  F^{(U)} \wedge F^{(U)} ~.
\end{align}


Next let us consider local rotation $U(x)^{ai}$ where the rotation matrix
depend only on the four-dimensional coordinates $x$. 
In this case, $U(x)^{ai}$ is not constant and the rotation cannot be
absorbed into gauge transformation. 
%
By using the quantities
\begin{align}
z^a(x) &= U^{ai}(x) y^i~,& W^a(x) &= U^{ai}(x) V^i~,&
U^{ai}(x) U^{bi}(x)  &= \delta^{ab}~,& U^{ci}(x) U^{cj}(x)  = \delta^{ij}~,
\end{align}
let us define a new gauge configuration
\begin{align}
A^{(U(x))} &:=  \frac{1}{ 4 \bq R_2} \gamma_{ab} z^{a}(x) W^{b}(x) ~.
\end{align}
This is not a solution of the equation of motion (\ref{eq:eom}) anymore. 
Rather one should consider it as deviation from 
the Cremmer-Scherk configuration.   
Although it can be split into background and fluctuation parts as 
in Eq.~(\ref{eq:fluc}), we keep it in the present form for our purpose. 
Because the derivatives $\partial / \partial x^{\mu}$ along the
four-dimensional space-time act on $U(x)$, 
the quantity $F^{(U(x))}$ has components 
along the four-dimensional space-time,
$F_{\rm (nc)}^{(U(x))}$: 
\begin{align}
F^{(U(x))} &= \frac{1}{ 4 \bq} 
 dx^{\mu} \left( \frac{\partial}{\partial x^{\mu}} U^{ac}(x) U^{bd}(x)
 \right) \gamma_{ab} y^{c}  \frac{dy^{d}}{(1+|y|^2/4)}  
+ \frac{1}{4 \bq R_2^2 }  \gamma_{ab}W^a(x) \wedge W^b(x) ~.
\end{align}
$F_{\rm (nc)}^{(U(x))}$ is a part of the field strength, which has
four-dimensional components,  
\begin{align}
F_{\rm (nc)}^{(U(x))}&:= 
\frac{1}{ 4 \bq} 
 dx^{\mu} \left( \frac{\partial}{\partial x^{\mu}} U^{ac}(x) U^{bd}(x)
 \right) \gamma_{ab} y^{c}  \frac{dy^{d}}{(1+|y|^2/4)}    \cr
&= \frac{1}{ 4 \bq R_2 } 
 dx^{\mu}  \alpha_{\mu}(x)^{ae} 
 \gamma_{ab} z^{e}(x) W^{b}(x) 
+  \frac{1}{ 4 \bq R_2 } 
 dx^{\mu} \alpha_{\mu}(x)^{bf} 
\gamma_{ab} z^{a}(x) W^{f}(x),  
\end{align}
where we have used the pull-back  of the
Maurer-Cartan form,  
\begin{align}
\alpha_{\mu}(x)^{ab}  &:=  \left( \frac{\partial}{\partial
  x^{\mu}} U^{ac}(x) \right)   {(U^{-1})}^{cb}(x)  ~.
\end{align}
The six-dimensional part $F_{\rm (c)}^{(U(x))}$ 
defined in Eq.~(\ref{eq:F,Fc,Fnc}) 
satisfies the  self-duality equation
\begin{align}
F_{\rm (c)}^{(U(x))} &:= \frac{1}{4 \bq R_2^2 }  \gamma_{ab}W^a(x) \wedge
W^b(x)~,&  
F_{\rm (c)}^{(U(x))}   &= *_6  \frac{\bq R_2^2}{3} \bi \gamma_7  F_{\rm (c)}^{(U(x))}
\wedge F_{\rm (c)}^{(U(x))} ~. 
\end{align}
At each fixed point $x$, this configuration is the same
as that in the previous  section. 
Let us substitute this into $S_{\rm (c)}$
\begin{align}
S_{\rm (c)}[ F_{\rm (c)}^{(U(x))} ]  &=  - \frac{1}{16}    \int dv^{(4)} \int {\rm Tr}
\left[ - F_{\rm (c)}^{(U(x))}  \wedge *_6 F_{\rm (c)}^{(U(x))} 
  \right. \cr
& \left. + \alpha^2 (
  F_{\rm (c)}^{(U(x))} \wedge F_{\rm (c)}^{(U(x))} )  \wedge *_6 (
  F_{\rm (c)}^{(U(x))} \wedge 
  F_{\rm (c)}^{(U(x))} )   \right]    \cr
&=    \bi \frac{1}{16}      \int dv^{(4)}  \frac{\bq R_2^2}{3}  
\left(1 +e^{-4 \phi_2} \right)  
 \int {\rm Tr} \left[ \gamma_7   F_{\rm (c)}^{(U(x))} \wedge
   F_{\rm (c)}^{(U(x))}\wedge F_{\rm (c)}^{(U(x))}     \right]   ~.
\label{eqn:act1}
\end{align}
Because of the self-duality property, the action becomes total
derivative as the integral over the six-dimensional sphere with the
same value that of $F^{(0)}$  
\begin{align}
 \int {\rm Tr} \left[ \gamma_7   F_{\rm (c)}^{(U(x))} \wedge
   F_{\rm (c)}^{(U(x))}\wedge F_{\rm (c)}^{(U(x))}    \right]   
&= \int {\rm Tr} \left[ \gamma_7   F^{(0)}_1 \wedge F^{(0)}_1\wedge
   F^{(0)}_1    \right] 
= \frac{96 \pi^3}{\bq^3} ~ \label{eq:value}
\end{align}
where we have used the fact that $U^{ab}(x)$ is a rotation matrix with unit
determinant, the relation ${\rm Tr} \gamma_7 \gamma_{ab}\gamma_{cd}\gamma_{ef} 
\sim \epsilon^{abcdef}$ which is an invariant tensor under rotations, 
and finally Eq.~(\ref{eq:charge1}). 
Of course, as an integral over four-dimensional space-time,
Eq.(\ref{eqn:act1}) is not a total  derivative. 

Let us define scalar fields $\pi(x)$ 
by $U(x) \equiv e^{ \pi (x)}$. 
A mass term of $\pi(x)$ in the low-energy effective theory on
 four-dimensional space-time could be found in $S_{\rm (c)}$ 
if it would exist.
Since $S_{\rm (c)}$ does not depend on $U(x)$  
we find that $\pi(x)$ are massless fields which are 
the candidate of the Nambu-Goldstone bosons.

Let us expand $\alpha_{\mu}(x)$ with respect to small fields $\pi(x)$. 
The leading
 term is just the derivative of $\pi$,  
\begin{align}
\alpha_{\mu}(x)  &=  \left( \frac{\partial}{\partial x^{\mu}} U (x)
\right)  U^T(x)  
=  \left( \frac{\partial}{\partial x^{\mu}}   e^{\pi (x)}  \right)
e^{- \pi (x)}  
=   \partial_{\mu} \pi (x)  + \cdots . 
\end{align}
Keeping in mind the facts that the action can be divided into two parts 
($S_{\rm (c)}$ and $S_{\rm (nc)}$),
and that $S_{\rm (c)}$ is independent of $U(x)$ or $\pi(x)$ 
as shown in Eqs.~(\ref{eqn:act1}) and (\ref{eq:value}), 
we obtain
\begin{align}
S[\pi(x)]  &= \frac{1}{16} \int {\rm Tr} \left\{  - F^{(U(x))} \wedge
* F^{(U(x))}  
+ \alpha^2 ( F^{(U(x))} \wedge F^{(U(x))} ) \wedge * ( F^{(U(x))}
\wedge F^{(U(x))} )  \right\} \cr 
&= \frac{1}{16} \int {\rm Tr} \left\{  - F_{\rm (nc)}^{(U(x))} \wedge *
F_{\rm (nc)}^{(U(x))}  \right. \cr 
& \left. + \alpha^2 ( F_{\rm (nc)}^{(U(x))} \wedge F_{\rm
  (c)}^{(U(x))} +  F_{\rm (c)}^{(U(x))} 
\wedge F_{\rm (nc)}^{(U(x))}  ) \wedge * ( F_{\rm (nc)}^{(U(x))}
\wedge F_{\rm (c)}^{(U(x))} + 
F_{\rm (c)}^{(U(x))} \wedge F_{\rm (nc)}^{(U(x))}   )  \right.\cr 
& \left. + \alpha^2 ( F_{\rm (nc)}^{(U(x))} \wedge F_{\rm
  (nc)}^{(U(x))} ) \wedge * ( 
F_{\rm (nc)}^{(U(x))} \wedge F_{\rm (nc)}^{(U(x))} )    \right\}  
   + \mbox{(terms~independent~of~ $\pi(x)$)}. 
\end{align}

Since in our notation gauge fields and field strength 
are written in terms of the gamma matrices 
let us rewrite $\alpha_{\mu}^{ab}$ 
in terms of the gamma matrices:  
$\alpha_{\mu}^{ab} \gamma_{ab}$.
%
%
By using this quantity,
$F_{\rm (nc)}^{(U(x))}$ can be rewritten as  
\begin{align}
F_{\rm (nc)}^{(U(x))} 
&=  \frac{1}{2} \left[A^{(0)} ,  \frac{1}{2} \alpha_{\mu}^{cd}
  \gamma_{cd} \right]  \wedge dx^{\mu}   ~.
\end{align}

\bigskip
So far we have considered only gauge fields 
in the compact space 
$A^{(0)} + \Phi = 
 \frac{1}{ 4 \bq R_2} \gamma_{ab}
z^{a}(x) W^{b}(x)$  
but not the one $v_{\mu}(x)$ in the four dimensional space-time, 
the (1,3) part.
Here we consider the total gauge field  
\begin{align}
A_{\rm H} &:= v_{\mu}(x) dx^{\mu} +   \frac{1}{ 4 \bq R_2} \gamma_{ab}
z^{a}(x) W^{b}(x) ~. 
\end{align}
By a gauge transformation, the Nambu-Goldstone modes are
absorbed into the vector fields $v_{\mu}$ as
\begin{align}
v_{\mu} & \rightarrow u_{\mu} := v_{\mu} + \frac{1}{2 \bq} \left(
\frac{1}{2} \alpha_{\mu}^{cd} \gamma_{cd} \right)~
\end{align}
with
\begin{align}
 u_{\mu}^{ab} 
&=\left\{ U \left[ U^{-1}  v_{\mu} U + \frac{1}{2 \bq}  
 U^{-1} \frac{\partial}{\partial x^{\mu}} U 
  \right] U^{-1} \right\}^{ab}.
\end{align}
By taking unitary gauge, we obtain
\begin{align}
{\rm Tr} W \wedge * W &=  {\rm Tr} W_u \wedge * W_u~,\cr
W_u &:= d u + \bq u \wedge u~.
\end{align}
$u_{\mu}^{ab}$ are massive Proca fields with the mass, given by 
\begin{align}
S_{\rm mass}  &= - \frac{19}{9} \frac{16 \pi^3}{15} L_c^4 \int  \sqrt{-g^{(4)}} d^4x 
 u_{\mu}^{ab}
u^{\mu,ab} ~.
\end{align}
This is nothing but the Higgs mechanism. 
Here we note that there are no cross terms between $u$ and
$\phi_2$ up to quadratic order and 
the whole mass matrix is block diagonal 
between $u$ and $\phi_2$.
The number of the Nambu-Goldstone fields 
$\pi$, fifteen, is the same as the 
dimension of Lie algebra so(6).   
Therefore, the gauge symmetry SO(6) is completely broken by the gauge
configuration.

\section{Conclusion}
\label{section:conclusion}

In this paper we have considered ten-dimensional Einstein-Yang-Mills theory
where the gauge field is given by Cremmer-Scherk configuration with a higher
derivative coupling on $S^6$. We have studied the consequences of 
ten-dimensional fluctuations of the gauge field on the stability of
the background metric and gauge field solutions and on the gauge
symmetry in the theory. 
The Cremmer-Scherk configuration is obtained by the identification of compact
direction and internal (gauge) direction as an extension of 't
Hooft-Polyakov monopole  \cite{Polyakov:1974ek}. 
By rotating the identification, 
we have obtained massless fluctuation modes 
identified as Nambu-Goldstone bosons \cite{Nambu:1961tp}. 
These massless modes are absorbed into vector fields and form massive
vector fields.  
Because there are fifteen Nambu-Goldstone modes, 
the gauge symmetry SO(6) is
completely broken and all the massless modes 
are absorbed into vector fields. 
By scaling the vector fields so that the coefficients of their
kinetic terms are canonically normalized, we found that  
the mass is proportional to the inverse of the
radius of the compact space.  
We conclude that there are neither tachyonic nor massless modes 
in the physical spectrum around the background and that 
the configuration is stable.

We must point out that in this paper we have not considered
perturbations of gravity and this remains as an important  
future problem. One possible extension of the present work 
is to consider other internal manifolds such as 
the projective space ${\mathbb C}P^3$ instead of $S^6$
using a nontrivial gauge configuration given in \cite{Kihara:2008zg}.

In the paper by Forgacs and Manton \cite{Forgacs:1979zs} 
they used an $S^2$ reduction with non-trivial background gauge 
field to $SU(2)$ Yang-Mills instantons on ${\mathbb R}^2\times S^2$, 
resulting in Abrikosov-Nielsen-Olesen vortex solution 
in the Abelian-Higgs model on ${\mathbb R}^2$. 
Our case of the $S^6$ compactification 
may relate higher dimensional solitons 
in pure Yang-Mills theory to 
topological solitons in Yang-Mills-Higgs system
in four or five dimensions \cite{Eto:2006pg} 
such as 
wall-vortex-monopole composites \cite{Isozumi:2004vg} or
instanton-vortex composites \cite{Eto:2004rz}.

\section*{Acknowledgments}
 We wish to thank Misao Sasaki,
 Ignacio Zaballa and Eoin \'{O} Colg\'{a}in for many useful
 discussions.    
H.K.~ would like to show his appreciation to Yutaka Hosotani and Katsushi
 Ito for their comments.  
The work of M.N.~is supported in part by Grant-in-Aid for Scientific
Research (No.~20740141) from the Ministry
of Education, Culture, Sports, Science and Technology-Japan. 
The work of H.K.~is supported by the Interactive Research Center of Science, 
Graduate School of Science and Engineering,
 Tokyo Institute of Technology. 

\appendix

\section{Non-existence of covarianly constant functions belonging 
to the adjoint representation
 on the Cremmer-Scherk configuration}
In this appendix, we study covariantly constant functions 
on Cremmer-Scherk gauge configuration on $S^6$. 
Let us consider the equation $D_0^{(6)} \varphi =0$ where 
$\varphi = \varphi^{ab} \gamma_{ab}$ and the gauge field $A^{(0)}$ is 
$\displaystyle \frac{1}{4 \bq R_2} y^a V^b \gamma_{ab}$. 
The equation $D_0^{(6)} \varphi =0$ is written as
\begin{align}
&\frac{\partial}{\partial y^b} \varphi + \frac{1}{4(1+|y|^2/4)}
 y^a  \left[ \gamma_{ab} , \varphi \right]  = 0 ~.
\label{eqn:ccf}
\end{align}
Let us take the contraction of $y^b$ and Eq.(\ref{eqn:ccf}). 
Because $\gamma_{ab}$ is antisytmmetric in indices $a,b$,
the equation simply becomes as,   
\begin{align}
y^b \frac{\partial}{\partial y^b} \varphi  &=0 ~,& 
\frac{\partial}{\partial r} \varphi &=0~,
\end{align}  
where the radial coordinate is defined as $r=|y|$. 
This shows that $\varphi$ does not depend on $r$ and
 depends only on angular coordinates $\theta_i$,
 $(i=1,2,\cdots, 5)$. 
The partial differential operators $\partial / \partial y^a$ 
is rewritten in terms of these coordinates, 
\begin{align}
\frac{\partial}{\partial y^a} &= 
\hat{y}^a \frac{\partial}{\partial r} + \frac{1}{r} L_a
\end{align}
where $L_a$ are linear combination of $\partial / \partial \theta_i$ 
and do not depend on $r$. 
The unit vector $\hat{y}^a = y^a/r$ only depends on those angles. 
Then Eq. (\ref{eqn:ccf}) becomes
\begin{align}
& L_b \varphi + \frac{r^2}{(1+r^2/4)}
\hat{y}^a [ \gamma_{ab} , \varphi ] = 0 ~. 
\label{eqn:ccf2}
\end{align}
The first term in Eq. (\ref{eqn:ccf2}) 
and $\hat{y}^a [ \gamma_{ab} , \varphi ]$ do not depend on $r$. 
Therefore, it turns out that $\varphi$ is constant and
commute with all $\gamma_{ab}$.  
This means that $\varphi \equiv 0$.


\end{document}